\documentclass[12pt]{article}
\usepackage{amsmath}
\usepackage{latexsym}
\usepackage{amssymb}
\input{BoxedEPS}

\SetRokickiEPSFSpecial  %% for dvips by Tom Rokicki, for VMS
\HideDisplacementBoxes

\numberwithin{equation}{section}
 \renewcommand{\theequation}{\thesection.\arabic{equation}}
 \def\appendixa{
 \vskip 1cm
 \noindent
 {\Large \bf Appendix A}
 \vskip 0.5cm

 \setcounter{equation}{0}
 \def\theequation{A.\arabic{equation}}
 }
 \def\appendixb{
 \vskip 1cm
 \noindent
 {\Large \bf Appendix B}
 \vskip 1cm
 \par
 \setcounter{equation}{0}
 \def\theequation{B.\arabic{equation}}
 }
 \def\appendixc{
 \vskip 1cm
 \noindent
 {\Large \bf Appendix C}
 \vskip 1cm
 \par
 \setcounter{equation}{0}
 \def\theequation{C.\arabic{equation}}
 }
 \def\appendixd{
 \vskip 1cm
 \noindent
 {\Large \bf Appendix D}
 \vskip 1cm
 \par
 \setcounter{equation}{0}
 \def\theequation{D.\arabic{equation}}
 }
\begin{document}

\title{Chern-Simons Modification of General Relativity}
\author{R. Jackiw$^1$ \negthickspace \footnote{Email: jackiw@lns.mit.edu}\enspace and S.-Y. Pi$^2$\footnote{Email: soyoung@buphy.bu.edu}\\
\\
{\small\itshape $^1$Center for Theoretical Physics}
  \\[-1ex]
{\small\itshape Department of Physics}\\[-1ex]
{\small\itshape Massachusetts Institute of
Technology} \\[-1ex]
{\small\itshape  Cambridge, Massachusetts 02139}\\
{\small\itshape $^2$Department of Physics}\\[-1ex]
{\small\itshape Boston University}\\[-1ex]
{\small\itshape Boston, Massachusetts 02215}}

\date{\small MIT-CTP-3409\\ BUHEP-03-18}

\maketitle

\pagestyle{myheadings} \markboth{R.Jackiw}{CS modified $GR$}
\thispagestyle{empty}

\begin{abstract}

General relativity is extended by promoting the three-dimensional gravitational Chern-Simons term to four dimensions. This entails choosing an embedding
coordinate $v_\mu$ -- an external quantity, which we fix to be a non-vanishing constant in its time component. The theory is identical to one in which the embedding coordinate is itself
a dynamical variable, rather than a fixed, external quantity. Consequently diffeomorphism symmetry breaking is hidden in the modified theory: the Schwarzschild metric is
a solution; gravitational waves possess two polarizations, each traveling at the velocity of light; a conserved energy-momentum (pseudo-) tensor can be constructed. The modification
is visible in the intensity of gravitational radiation: the two polarizations of a gravity wave carry intensities that are suppressed/enchanced by the extension.

\end{abstract}

\section{Introduction}

Although the Chern-Simons term was first introduced in three-dimensional gauge field and gravitational
models \cite{rj1}, it can also deform physical theories in four-dimensional space-time, where it modifies conventional
kinematics and dynamics in a Lorentz and CTP violating fashion. This possibility has been investigated for Maxwell
electrodynamics. \cite{rj2} In the present paper we study a similar deformation of  Einstein's general relativistic gravity
theory.

To set the stage for our analysis, we review the electromagnetic example. In Chern-Simons modified Maxwell theory,
the homogenous Maxwell equation for the dual electromagnetic field

\begin{equation}
^\ast F^{\mu \nu} \equiv \frac{1}{2} \varepsilon^{\mu \nu \alpha \beta} F_{\alpha \beta}
\end{equation}
is retained (``Bianchi identity")

\begin{equation}
\partial_\mu {^\ast F}^{\mu \nu}=0,
\label{bianchi}
\end{equation}
so that we still have the possibility of introducing potentials $A_\mu$.

\begin{equation}
F_{\mu \nu} = \partial_\mu A_\nu -\partial_\nu A_\mu
\end{equation}
But the inhomogeneous equation, $\partial_\mu F^{\mu \nu} = J^\nu$, is extended to

\begin{equation}
\partial_\mu F^{\mu \nu} + v_\mu {^*F}^{\mu \nu} = J^\nu.
\label{gaugevarience}
\end{equation}
While gauge invariance holds, Lorentz invariance is lost owing to the presence of an external, fixed four-vector $v_\mu$, called the ``embedding coordinate." 
Taking $v_\mu$ to be time-like, and in its rest frame: $v_\mu = (\mu, \bf 0)$, preserves rotational symmetry but not Lorentz boost invariance. Only Amp\`{e}re's law is modified.

\begin{eqnarray}
-\frac{\partial}{\partial t}& \negthickspace {\bf E} + \nabla {\bf \times} \ {\bf B} = &{\bf J} + \mu {\bf B}\\ \nonumber
&E^i \equiv F^{i0},&B^i \equiv -\frac{1}{2} \varepsilon^{ijk} F_{jk} 
\label{1.5}
\end{eqnarray}
Note that current conservation is still the only consistency requirement on the extended equation (1.4), since the left side remains divergence-free by virtue of ~(\ref{bianchi}).

The kinetic portion  [left side of ~(\ref{gaugevarience})] is obtained by varying $A_\mu$ in a Chern-Simons modified Maxwell action.

\begin{subequations}
\begin{equation}
I= \int d^4  x \left(-\frac{1}{4} \ F^{\mu \nu} F_{\mu \nu} + \frac{1}{2} \  v_\mu {^*F}^{\mu \nu} A_\nu \right)
\label{1.6a}
\end{equation}

\begin{equation}
I = \int d^4  x \left(- \frac{1}{4} \ F^{\mu \nu} \,F_{\mu \nu} + \frac{\mu}{2} \ \bf A \cdot \bf B \right)
\label{1.6b}
\end{equation}
\end{subequations}
In the second formula we have used the special, time-like form for $v_\mu$, and the modification of the action involves in the (Abelian)
Chern-Simons expression $\bf A \cdot \bf B$, \cite{rj1} while in~(\ref{1.6a}) there appears the (Abelian) Chern-Simons current $ 
{^*F^{\mu \nu}} A_\nu$, whose divergence gives the topological Pontryagin density.
\begin{equation}
 \partial_\mu \left(  {^*F}^{\mu \nu} A_\nu \right) = \frac{1}{2} \,{^*F}^{\mu \nu} F_{\mu \nu} \equiv \frac{1}{2} {^*FF}
\label{1.7}
\end{equation}
Note that the extended Lagrange densities in (\ref{1.5}) are not gauge invariant, even though the equations of motion possess this property.
Correspondingly, eq.~(\ref{1.7}) shows that an equivalent but gauge invariant Lagrange density 
can be obtained by adjusting ~(\ref{1.6a}) by a total derivative.

\begin{equation}
{\rm I}= -\frac{1}{4} \int d^4  x \left (F^{\mu \nu} F_{\mu \nu} + \theta \ {^*F} F \right)
\label{1.8}
\end{equation}
Now the external, fixed quantity is $\theta$, which is set equal to $v_\mu x^\mu$ or $\mu t$, so that ~(\ref{1.6a}) or ~(\ref{1.6b}) are reproduced.

The physical consequences of the modified theory are the following. Owing to gauge invariance, the photon retains only two polarizations, but in vacuum they 
travel with velocities that differ from c (Lorentz boost invariance is lost) and from each other (parity invariance is lost). For example, with plane monochromatic waves, which
continue to solve the modified equations, the dispersion law between frequency $\omega$ and wave number $\left| \bf k \right|$ is changed to $\omega = \left| \bf k \right| 
\pm \mu /2$, for small $\mu$. (For arbitrary $\mu$, the exact dispersion law shows an instability for small 
$\left| \bf k \right|$: $\omega= \sqrt {\left| \bf k \right|^2 \pm \mu \left| \bf k \right|})$. This birefringence of the vacuum produces a Faraday-like rotation on polarized light. Actual
measurements of light from distant galaxies have shown that no such extension of 
electromagnetism exists in Nature. \cite{rj1, rj2}

Carrying out similar constructions for gravity theory requires first deciding how to embed the three-dimensional gravitational Chern-Simons term into four-dimensional
general relativity. We choose to begin with the gravity analog to~(\ref{1.8}). As we shall demonstrate, this leads to a theory where symmetry breaking effects are hardly visible;
they are suppressed by dynamics. The Schwarzschild solution still holds.
Gravity waves still possess two polarizations, which propagate with velocity c. But
parity violation manifests itself in that the two polarizations carry different intensities. A symmetric, divergenceless, second rank (pseudo-) tensor survives
in the extended theory; it can serve as the gravitational energy-momentum (pseudo-) tensor. 

Our paper is organized as follows.
In the next Section we present the modified gravity theory and discuss a novel consistency condition on the modified equations of motion. In Section 3, a stationary {\it Ansatz} is made and the Schwarzschild solution is regained
in the spherically symmetric case. Section 4 is devoted to the linear approximation, wherein one exhibits propagating physical degrees of freedom, finds gravitational waves and
identifies a conserved, symmetric gravitational energy-momentum (pseudo-) tensor. A concluding statement comprises the last Section 5. Details of technical computations are relegated to Appendices.

\section{Chern-Simons Modified Gravity}

The gravitational Chern-Simons term CS$(\Gamma)$ is the three-dimensional quantity [1]
\begin{equation}
CS(\Gamma) = \frac{1}{4 \pi^2} \int d^3 x \  \varepsilon^{ijk} \ (\frac{1}{2}
\ ^3\Gamma^p_{i q} \, \partial_j  {^3 \Gamma^q_{k p}} + \frac{1}{3}  {^3\Gamma}^p_{iq} \, {^3\Gamma}^q_{j r} \, ^3\Gamma^r_{kp}).
\label{2.1}
\end{equation}
[Definitions of geometrical quantities follow Weinberg, \cite{rj4} except our metric tensor is of opposite sign:  $(1, -1,...)$. Latin letters range over three values, indexing coordinates on a three manifold; 
Greek letters denote analogous quantities in four dimensions; the superscript ``3" denotes three-dimensional objects.]  
The three-dimensional Christoffel connection is constructed in the usual way from the metric tensor, which is taken to be the fundamental dynamical variable. 
The variation of ~(\ref{2.1}) produces the Cotton tensor $^3C^{mn}$.

\begin{eqnarray}
\delta CS(\Gamma) \negthickspace \negthickspace \negthickspace \negthickspace &\quad =\frac{1}{4 \pi{^2}} \int d^3x \, \varepsilon^{ijm} \ {^3D_i}  {^3R^n_j}  \delta g_{mn}\\
&\equiv -\frac{1}{4 \pi{^2}} \int d^3 x \sqrt{g}~\ {^3C}^{mn} \delta g_{mn} \nonumber
\label{2.2}
\end{eqnarray}
Here $^3R^m_n$ is the three-dimensional Ricci tensor;  $^3D_i$ effects three-dimensional, covariant differentiation; $g$ is the metric determinant, context fixes
whether it is the three- or four-dimensional quantity. Eq. (2.2) implies

\begin{equation}
^3C^{mn} = -\frac{1}{2 \sqrt{g}} \left (\varepsilon^{mij} \, {^3D_i} {^3R^n_j} + \varepsilon^{nij} \,{^3D_i}
{^3R^m_j} \right).
\label{2.3}
\end{equation} 
$^3C^{mn}$ is symmetric, traceless and covariantly conserved (in the three-dimensional sense).

A related four-dimensional quantity is the Chern-Simons topological current
\begin{equation}
K^\mu = 2 \varepsilon^{\mu}  {^{\alpha \beta \gamma}} \left [\frac{1}{2} ~ \Gamma^\sigma_{\alpha \tau} \, \partial_ \beta
~ \Gamma^\tau_{\gamma \sigma} + \frac{1}{3} ~  \Gamma^\sigma_{\alpha \tau} \,  \Gamma^\tau_{\beta \eta} \Gamma^\eta_{\gamma \sigma} \right],
\label{2.4}
\end{equation}
which satisfies
\begin{equation}
\partial_\mu K^\mu = \frac{1}{2} {^*R}^\sigma_{~\tau} \ {^{\mu \nu}} \ R^\tau {_{\sigma \mu \nu}} \equiv \frac{1}{2} {^*RR},
\label{2.5}
\end{equation}
where $R^\tau_{~ {\sigma \mu \nu}}$ is the four-dimensional Riemann tensor
\begin{eqnarray}
R^\tau_{~ {\sigma \mu \nu}} = &\partial_\nu ~ \Gamma_{\mu \sigma} ^\tau  - \partial_\mu ~ \Gamma_{\nu \sigma} ^\tau \negthickspace \negthickspace &+ \Gamma^\tau_{\nu \eta} \Gamma^\eta_{\mu \sigma}
 - \Gamma^\tau_{\mu \eta} \Gamma^\eta_{\nu \sigma},
\label{2.6} 
\end{eqnarray}
and $^\ast R^{\tau \ \mu \nu}_{\, \,\sigma}$ is its dual
\begin{equation}
^*R^{\tau \ \mu \nu}_{\, \,\sigma} = \frac{1}{2} ~ \varepsilon^{\mu \nu \alpha \beta} R^\tau_{~{\sigma \alpha \beta}}.
\label{2.7}
\end{equation}
Evidently the divergence of the topological current is the gravitational Pontryagin density in analogy with the electromagnetic case [see ~(\ref{1.7})]. Note however
that $K^0$ is not related to the Chern-Simons term ~(\ref{2.1}): the former involves four-dimensional Christoffel connections [even though some of the indices $\alpha, \beta, \gamma$
are spatial at $\mu=0$], while in the latter only three-dimensional Christoffel connections are present. Because of this difference, there are different
possibilities for extending the Einstein theory by a Chern-Simons term.

We choose, again by analogy to electromagnetism [see (\ref{1.8})], the following expression as the extension of the
Hilbert-Einstein action,
\begin{equation}
{\rm I}= \frac{1}{16 \pi G} \int d^4 x \left( \sqrt{-g} R + \frac{1}{4} \theta {^*RR} \right) = \frac{1}{16 \pi G} \int d^4 x \left(\sqrt{-g} R - \frac{1}{2} v_\mu K^\mu \right)
\label{2.8}
\end{equation}
where $\theta$ is a prescribed external quantity, and $v_\mu \equiv \partial_\mu \theta$ is the embedding coordinate. The second equality follows with the help of formula (\ref{2.5}). The variation of the first term in the integrand with respect to $g_{\mu \nu}$ produces the usual Einstein tensor 
$G^{\mu \nu} \equiv R^{\mu \nu} -\frac{1}{2} ~ g^{\mu \nu} ~ R \  (R_{\mu \nu} \equiv R^{\tau}_{~\mu \tau \nu}, \ R \equiv R^\mu_\mu)$. The variation of the second, topological term gives a traceless symmetric, second-rank tensor, which we name the four-dimensional Cotton tensor $C^{\mu \nu}$.

\begin{equation}
\delta I_{CS} =\delta \frac{1}{4}  \int \negthinspace d^4 x \,   \theta {^*RR} \equiv \int d^4 x \sqrt{-g} C^{\mu \nu} \delta g_{\mu \nu} = -\int \negthinspace \negthinspace
\begin{array}{l}
d^4 x \sqrt{-g}
  \displaystyle C_{\mu \nu}  \delta g^{\mu \nu}
\end{array}
\label{2.9}
\end{equation}

\begin{eqnarray}
C^{\mu \nu}=-\frac{1}{2 \sqrt{-g}} \left[ v_\sigma\left(\varepsilon^{\sigma \mu \alpha \beta} D_\alpha R^\nu_\beta + \varepsilon^{\sigma \nu \alpha \beta}
D_\alpha R^\mu_\beta \right) + v_{\sigma \tau} \left(^*R^{\tau \mu \sigma \nu} + ^*R^{\tau \nu \sigma \mu} \right) \right]
\label{2.10}
\end{eqnarray}
Formula (\ref{2.10}) is derived in Appendix A. Here $v_{\sigma \tau}$ is the covariant derivative of the embedding coordinate,   $v_{\sigma \tau} \equiv D_\sigma v_\tau = D_\sigma D_\tau \theta$. When $\theta$ is taken to be linear in $x$, $\theta = x^\sigma v_{\sigma}$,$~v_\sigma$ is constant and  $v_{\sigma \tau} = - \Gamma^\alpha_{\sigma \tau} v_\alpha$. More specifically for $v_\sigma = (1/\mu, \mathbf 0)$ the first contribution to $C^{\mu \nu}$ is
similar to the three-dimensional Cotton tensor ~(\ref{2.3}), except that four-dimensional geometric entities are now present. Moreover, evaluated on a stationary metric tensor $\sqrt{-g} C^{mn}$ coincides with $\sqrt{g} \,^3C^{mn}$ (see the next Section).
[To achieve this coincidence the factor $\frac{1}{4}$ is introduced in $I_{CS}$].  Even the second term in (\ref{2.10}) can be understood as a four-dimensional generalization: the (dual of the) Riemann tensor, which occurs in (\ref{2.10}), can be written in terms of the Weyl conformal tensor, the Ricci tensor and scalar; the Ricci quantities fail to contribute, so that
the second term in (\ref{2.10}) involves just the (dual) Weyl tensor. However, in three dimensions the Weyl tensor identically vanishes, and that is why it does not appear in the
three-dimensional Cotton tensor (\ref{2.3}).

(Actually, Cotton defined his tensor purely geometrically for arbitrary dimensions $(d)$ as $D_\alpha \tilde{R}^\mu_\beta - D_\beta \tilde{R}^\mu_\alpha, \ \tilde{R}^\mu_\alpha \equiv R^\mu_\alpha - 
\frac{1}{2(d-1)} \delta^\mu_\alpha R$. \cite{rj5} For $d=3$, this is equivalent to our (\ref{2.3}), and so can also be given a variational definition, as in (2.2), (\ref{2.3}).
In other dimensions, Cotton's formula does not appear to possess a variational formulation, whereas our formula (\ref{2.9}), (\ref{2.10}) does -- at the expense of having introduced
the non geometric, external entites $\theta, v_\mu$.) 

The proposed deformation of Einstein's general relativity equation reads
\begin{equation}
G^{\mu \nu} + C^{\mu \nu} = -8 \pi G T^{\mu \nu}.
\label{2.11}
\end{equation}
It is necessary to consider the consistency condition that follows upon taking the covariant divergence of (\ref{2.11}).
The Bianchi identity enforces $D_\mu G^{\mu \nu}=0$, while diffeomorphism invariant dynamics for the matter degrees of freedom implies that the energy-momentum
tensor $T^{\mu \nu}$ similarly satisfies $D_\mu T^{\mu \nu} =0$.
However, the covariant divergence of the four-dimensional Cotton tensor is non-zero, in contrast to the situation in three dimensions. The divergence of (\ref{2.10}),
evaluated in Appendix B, is found to be
\begin{equation}
D_\mu C^{\mu \nu} = \frac{1}{8 \sqrt{-g}} v^\nu {^*RR}.
\label{2.12}
\end{equation}
Thus the equations of the extended theory ~(\ref{2.11}) possess solutions that are necessarily confined to spaces with vanishing $^*RR = 2 \partial_\mu K^\mu$. 

One may also understand this condition by studying the response of ${\rm I}_{CS}=\frac{1}{4} \int d^4  x \  \theta {^*RR}$ to an arbitrary, infinitesimal coordinate transformation
\begin{equation}
\delta x^\mu = -f^\mu (x).
\label{2.13}
\end{equation}
Because $^*RR$ is a coordinate density, it responds to ~(\ref{2.13}) as $\delta (^*RR) = \partial_\mu (f^\mu ~{^*RR})$, and $\theta$, being an external parameter, does not transform.
\begin{equation}
\delta {\rm I}_{CS} = \frac{1}{4} \int d^4  x \ \theta \ \partial_\mu (f^\mu {^*RR}) = -\frac{1}{4} \int d^4  x \ v_\mu \ f^\mu {^*RR}
\label{2.14}
\end{equation}
Alternatively, we can evaluate $\delta {\rm I}_{CS}$ by explicitly varying $g_{\mu \nu}$ according to
\begin{equation}
\delta g_{\mu \nu} = D_\mu f_\nu + D_\nu f_\mu,
\label{2.15}
\end{equation}
and from ~(\ref{2.9}) we have
\begin{equation}
\delta {\rm I}_{CS} = 2 \int d^4 x \sqrt{-g}~\ C^{\mu \nu} D_\mu f_\nu = -2 \int d^4  x  \sqrt{-g}  \ ~{(D_\mu C^{\mu \nu})} f_\nu.
\label{2.16}
\end{equation}
Comparison with ~(\ref{2.14}) regains (\ref{2.12}), since $f^\mu$ is arbitrary.

Thus the Cotton tensor's non-vanishing divergence, proportional to ${^\ast RR}$, is a measure of the failure of diffeomorphism invariance when $v_\mu$ is non-vanishing.
But consistency of dynamics forces $^\ast RR$ to vanish, so in a sense diffeomorphism symmetry breaking is suppressed dynamically.

For another perspective on this, consider an extended theory, where $\theta$ is taken to be a local dynamical variable, acting as a Lagrange multiplier for $^\ast RR$.
The Chern-Simons action is now invariant, because $\theta$, being dynamical, now transforms under ~(\ref{2.13}) as a scalar: $\delta \theta = f^\mu \partial_\mu \theta = f^\mu v_\mu $ .
Consequently ~(\ref{2.14}) acquires an additional contribution $\frac{1}{4} \int d^4 x \, \delta \theta ~{^\ast RR} = \frac{1}{4} \int d^4 x f^\mu v_\mu
{^\ast RR}$, which cancels the nonvanishing result in ~(\ref{2.14}) and shows that ${\rm I}_{CS}$ is invariant. However, the equation of motion arising from $g_{\mu \nu}$
variation remains as ~(\ref{2.11}), while variation of the dynamical $\theta$ forces $^\ast RR$ to vanish -- a requirement which is already enforced by ~(\ref{2.11}).
So the manifestly invariant theory, with dynamical $\theta$, possesses the same equations of motion as the theory with external $\theta$.
In other words, in the fully dynamical and coordinate invariant theory, $\theta$ as well as the embedding coordinate $v_\sigma \equiv \partial_\sigma \theta$, take values
that are arbitrary, as long as the modified gravity equation (\ref{2.11}) possesses non trivial solutions. On the other hand, in our extended, non invariant gravity theory, the embedding
coordinate takes a fixed value $({\it e.g.} \, v_\sigma = \frac{1}{\mu} \delta_{\sigma 0})$, which still supports non trivial solutions to (\ref{2.11}). 

[Note that this mechanism for dynamical suppression of diffeomorphism symmetry violation is not confined to the Chern-Simons modification considered above. Indeed any contribution to the gravity action of the form 
${\rm I}_X = \int d^4 x ~\theta X$, where X is a coordinate scalar density, and $\delta {\rm I}_X \equiv \int d^4 x \sqrt{-g} ~X^{\mu \nu} \delta g_{\mu \nu}$, will
lead to the addition of $X^{\mu \nu}$ in the Einstein equation. Consistency of the modified equation demands that $D_\mu X^{\mu \nu} = 0$. Moreover, an argument as in ~(\ref{2.13}) -- ~(\ref{2.16}) shows that  $D_\mu X^{\mu \nu} = \frac{1}{2 \sqrt{-g}} v^\nu X$.
Thus the consistency condition becomes $X=0$ and it is equivalently enforced by varying ${\rm I}_X$ with respect to $\theta$. This feature is absent in non gravitational theories: for example,
varying $\theta$ in Chern-Simons extended Maxwell theory produces $^*FF=0$ [see (\ref{1.8})] -- a condition that does not follow from the equation of motion (\ref{gaugevarience})]

Eq. (\ref{2.14}) also shows that with $v_\mu = \left(\frac{1}{\mu}, \bf 0 \right)$, coordinate transformations in which time is transformed only by shifting its origin, {\it i.e.} $f^0 = constant, f^i \, arbitrary$, leave $I_{CS}$ invariant.
Space-time dependent reparameterization of the spatial variables and time translation remain  conventional symmetries for the action of the extended theory. Henceforth, $v_\mu$ will be always taken at that time-like value.

\section{Persistence of Schwarzschild Solution}

We show that the modified equation (\ref{2.11}), where $C^{\mu \nu}$ is given by (\ref{2.10}), with $ v_\sigma =
 \frac{1}{\mu} \delta_{\sigma 0}, \ v_{\sigma \tau} = -\frac{1}{\mu} \,\Gamma^0_{\sigma \tau}$ and vanishing $T^{\mu \nu}$,
continues to support the Schwarzschild solution. Geometric quantities
are evaluated first for a metric tensor in stationary form
\begin{equation}
g_{\mu \nu} = \left(
\begin{array}{cc}
N& 0\\
0& g_{ij}
\end{array} \right)
\label{3.1}
\end{equation}
with time-independent entries, and then $g_{ij}$ is taken to be spherically symmetric.

\subsection*{Stationary Space-Time}

With ~(\ref{3.1}) the only non-vanishing Christoffel connection components are
\begin{equation}
\Gamma^0_{0i} = \Gamma^0_{i0} = \partial_i ~\ell n~ \sqrt{N}, \ \ \Gamma^i_{0  0} = -\frac{1}{2}~\ g^{ij} \partial_j N, \ \ \Gamma^m_{ij} = 
{^3\Gamma^m_{ij}}.
\label{3.2}
\end{equation}
The non vanishing components of the (dual) Riemann tensor are
\begin{subequations}
\begin{equation}
^\ast R^{0imn} = \varepsilon^{mnj} N^{i}_{j}
\label{3.3a}
\end{equation}
\begin{equation}
^\ast R^{ij0k} = \frac{1}{2} ~\varepsilon^{kmn}~ ^3R^{ij}_{~ mn} = -\varepsilon^{ij \ell}~^3G^k_\ell
\label{3.3b}
\end{equation}
\end{subequations}
(and partners with exchanged indices). In (\ref{3.3a}) we have defined
\begin{equation}
N^i_j = \frac{1}{\sqrt{N}} ~^3D^i ~^3D_j ~\sqrt{N},
\label{3.4}
\end{equation}
and in the second equality of (\ref{3.3b}), the three-dimensional Riemann tensor is expressed in terms of Ricci tensor and scalar, leading in
the end to the Einstein tensor.
\begin{equation}
^3R^{ij} \, {_{mn}} = \delta^i _m \, ^3G^j_n - \delta^i_n \, ^3G^j_m + \delta^j_n  \, ^3R^i_m - \delta^j_m \, ^3R^i_n
\end{equation}
Finally the non-zero Ricci tensor components read
\begin{subequations}
\begin{equation}
R^0_0 = N^i_i
\label{3.6a}
\end{equation}
\begin{equation}
R^i_j = N^i_j + {^3R^i_j}
\label{3.6b}
\end{equation}
\end{subequations}

According to ~(\ref{2.10}) and ~(\ref{3.2}), the stationary Cotton tensor becomes

\begin{eqnarray}
C^{\mu \nu} =  -\frac{1}{2 \mu \sqrt{-g}} \bigg[ \varepsilon^{0 \mu \alpha \beta} D_\alpha R^\nu_\beta + \varepsilon^{0 \nu \alpha \beta} D_\alpha R^\mu_\beta \nonumber\\
-\Gamma^0_{0i} \left ({^*R}^{i \mu 0 \nu} + {^*R}^{i \nu 0 \mu} + {^*R}^{0 \mu i \nu} + {^*R}^{o \nu i \mu} \right) \bigg].
\label{3.7}
\end{eqnarray}
We see from (3.3)  and ~(\ref{3.7}) that $C^{00}$ and $C^{0n}=C^{n0}$ vanish. There remains the space-space component.
\begin{eqnarray}
C^{mn} = -\frac{1}{2 \mu \sqrt{-g}} \bigg(
\varepsilon^{mij} D_i R^n_j + \varepsilon^{nij} D_i R^m_j \nonumber\\ 
- \partial_i ~\ell n ~\sqrt{N} \left[{^\ast R^{im0n}} + {^\ast R}^{in0m} + {^*R}^{0min} + {^\ast R}^{0nim} \right] \bigg) \nonumber\\
=-\frac{1}{2 \mu \sqrt{-g}} \bigg(\varepsilon^{mij} ~ ^3D_i ~ ^3R^n_j + \varepsilon^{nij} ~ {^3D_i} ~^3R^m_j \bigg) \nonumber\\
 -\frac{1}{2 \mu \sqrt{-g}} \bigg (\varepsilon^{mij} \bigg[{^3D}_i N^n_j + \partial_i \ell n \sqrt{N} ~(N^n_j -{^3G^n_J}) \bigg] \nonumber\\
 +\varepsilon^{nij} \bigg[{^3D}_i N^m_j + \partial_i \ell n ~(N^m_j -{^3G^m_J}) \bigg] \bigg)
\label{3.8}
\end{eqnarray}
We show in Appendix C that the bracketed expressions in the second equality vanish, leaving the first term as the sole contribution to the stationary, four-dimensional Cotton tensor. Note that 
it coincides with the three-dimensional Cotton tensor in ~(\ref{2.3}) apart from normalization.

\subsection*{Spherically Symmetric Space}
We evaluate the stationary [three-dimensional: ($r, \theta, \varphi$)] Cotton tensor (\ref{2.3}) on a spherically symmetric space-time, with non-vanishing metric tensor components
\begin{eqnarray}
g_{rr} = 1/g^{rr} = -A(r) \nonumber\\
g_{\theta \theta} = 1/g^{\theta \theta} = -r^2 \nonumber\\
g_{\varphi \varphi} = 1/g^{\varphi \varphi} = -r^2 \sin^2 \theta.
\end{eqnarray}
(The calculation is completely three-dimensional, and the superscript ``3" is omitted.) The expressions for the Christoffel connection and the Ricci tensor are familiar.
\cite{rj4} The non vanishing components of the former are
\begin{equation}
\begin{array}{l}
\Gamma^r_{rr} = A^{'} /2 A, ~\Gamma^r_{\varphi \varphi} = \sin^2 \theta \Gamma^r_{\theta \theta} = - r \sin^2 ~\theta / A\\
\Gamma^\theta_{\theta r} = \Gamma^\theta_{r\theta} = \Gamma^\varphi_{\varphi r} = \Gamma^\varphi_{r \varphi} = 1/r,\\
\Gamma^\theta_{\varphi \varphi} = -sin^2 \ \theta \ \Gamma^\varphi_{\varphi \theta} = -sin^2 \  \theta \ \Gamma^\varphi_{\theta \varphi} = -\sin \theta \cos \theta.
\end{array}
\label{3.10}
\end{equation}
The latter possess only diagonal components
\begin{equation}
\begin{array}{l}
R^r_r = A^{'}/ rA^2\\
R^\theta_\theta = R^\varphi_\varphi = \frac{1}{r^2} (1-1/A) + A^{'}/2rA^2.
\end{array}
\label{3.11}
\end{equation}
(Differentiation with respect to $r$ is denoted by the dash.) Evaluation of \\ $\epsilon^{mij} D_iR^n_j = \epsilon^{mij} (\partial_i R^n_j + \Gamma^n_{ik} R^k_j)$  shows that $C^{mn}$ vanishes in all components.

Consequently the Schwarzschild solution holds in the extended theory, which therefore passes the three classic tests of general relativity.

Evidently the Schwarzschild metric leads to vanishing $^*RR$, since the modified gravity equations (\ref{2.11}) are satisfied. Correspondingly the Kerr metric, for which ${^*RR} \ne 0$, cannot be a solution.
It is an outstanding open question which deformation of the Kerr solution solves the deformed gravity equations.

\section{Linear Theory}

We now analyze the extended equations by linearizing the metric $g_{\mu \nu}$ around a Minkowski background metric $\eta_{\mu \nu}$.

The left side of ~(\ref{2.11}) in linear approximation $g_{\mu \nu} = \eta_{\mu \nu} + h_{\mu \nu}$ is denoted by $G^{linear}_{\mu \nu} + C^{linear}_{\mu \nu}$, where 
\begin{equation}
\begin{array}{lll}
G^{linear}_{\mu \nu}&=\frac{1}{2} (\Box h_{\mu \nu} + \partial_{\mu} \partial_{\nu} h - \partial_\mu ~\partial_{\alpha} h^\alpha_\nu - \partial_\nu \partial_\alpha h^\alpha_\mu 
 -\eta_{\mu \nu} [\Box h - \partial_\alpha ~\partial_\beta ~h^{\alpha \beta}]),\\ [10pt]

C^{linear}_{\mu \nu}&=-\frac{1}{4 \mu} (\varepsilon_{0 \mu \alpha \beta} \, \partial^\alpha [\Box h_ \nu^\beta - \partial_\nu \,\partial_\gamma h^{\gamma \beta}]
 + \varepsilon_{0 \nu \alpha \beta} ~\partial^\alpha [ \Box h^\beta_\mu - \partial_\mu \partial_\gamma h^{\gamma \beta}]). 
\end{array} 
\label{4.1} 
\end{equation}
\noindent
Here $h=h^\mu_\mu$ and indices are moved by $\eta^{\mu \nu}$. Since $D^\mu G_{\mu \nu} = 0$, the linear portion reads $ \partial^\mu G^{linear}_{\mu \nu} = 0$. The covariant divergence
of $C_{\mu \nu}$ is not zero, but it is quadratic in $h_{\mu \nu}$, hence the linear part of $C_{\mu \nu}$ also satisfies $\partial^\mu C^{linear}_{\mu \nu} = 0$. Both statements can be checked explicitly
from ~(\ref{4.1}). One verifies that both $G^{linear}_{\mu \nu}$ and $C^{linear}_{\mu \nu}$ are invariant against the ``gauge transformation"
\begin{equation}
h_{\mu \nu} \to h_{\mu \nu} + \partial_\mu \lambda_\nu + \partial_\nu \lambda_\mu.
\label{4.2}
\end{equation}
The linear equation of motion follows from the quadratic portion of the action ~(\ref{2.8}), which reads
\begin{equation}
{\rm I}_{quadratic} = -\frac{1}{2} \int d^4 x\ h^{\mu \nu} \ \bigg(G^{linear}_{\mu \nu} + C^{linear}_{\mu \nu} \bigg).
\label{4.3}
\end{equation}

\subsection*{Degrees of Freedom}
The quadratic action allows determining the propagating degrees of freedom. To this end, we decompose $h^{\mu \nu}$ as follows
\begin{equation}
\begin{array}{lll}
h^{00} = n, \ h^{0i} = n^i_T + \partial_i n_L ,\\ [10 pt]
h^{ij} = \left(\delta^{ij} - \frac{\partial_i \partial_j}{\nabla^2} \right) \varphi + \frac{\partial_i \partial_j}{\nabla^2} \chi + \left(\partial_i \ \xi^{j}_T + \partial_j \xi^{i}_T \right)
 + h^{ij}_{TT}.
\end{array}
\label{4.4}
\end{equation}
The subscript $T$ denotes spatial transversality; $TT$ additionally indicates spatial tracelessness. In three spatial dimensions the symmetric $h^{ij}_{TT}$ possesses two components.
Observe that the gauge transformation ~(\ref{4.2})  does not affect $\varphi$ nor $h^{ij}_{TT}$. While the remaining components undergo a non-trivial response, the following combinations
are gauge invariant.
\begin{equation}
\begin{array}{ll}
\Lambda = \nabla^2 (n + 2 \dot{n}_L) + \ddot{\chi}\\ [10pt]
{\boldsymbol \sigma}_T = {\bf n}_T + \dot{\boldsymbol \xi}_T
\end{array}
\label{4.5}
\end{equation}
(The over dot denotes differentiation with respect to time.) With  ~(\ref{4.4}) and ~(\ref{4.5}), the quadratic action ~(\ref{4.3}) becomes
\begin{equation}
\begin{array}{ll}
I_{quadratic}& = \int d^4 \thinspace x \left[-\frac{1}{4} h^{ij}_{TT} \thinspace \Box h^{i j}_{TT} + \frac{1}{2} \varphi \Box \varphi
+ \frac{1}{2} (\partial_i \sigma^j_T)^2 + \varphi \Lambda \right]\\ [10 pt]
& +\frac{1}{4 \mu} \int d^4 x \thinspace \varepsilon^{ijk} \left[ h^{i \ell}_{TT} \partial_k \Box h^{\ell j}_{TT} + \sigma^i_T \partial_k \nabla^2 \sigma^j_T \right].
\end{array}
\label{4.6}
\end{equation}
In the absence of the $0(\frac{1}{\mu})$ Chern-Simons extension, the first Einstein-Hilbert term shows that $h^{ij}_{TT}$ is the only propagating component, $\varphi$ vanishes
by virtue of the Lagrange multiplier $\Lambda$, while ${\boldsymbol \sigma}_T$ does not propagate since only spatial derivatives act on it. This is not changed when the extension is included. So in empty space the
propagation of the gravitational field is still governed by the d'Alembertian acting on $h^{ij}_{TT}$,
{\it i.e.} there are two linearly independent polarizations, propagating as waves with velocity c, just as in the absence of the extension. For example, for monochromatic plane waves with
\begin{equation}
h^{\mu \nu}_{TT} = \varepsilon^{\mu \nu}_{TT} e^{i(\omega t - {\bf k} \cdot {\bf r})}, \qquad\ \Box h^{\mu \nu}_{TT} = 0,
\label{4.7}
\end{equation}
we have $\omega = |{\bf k}|$. Moreover, one also verifies that for these solutions the 
symmetry breaking quantity $^*RR$ vanishes in quadratic
order (see Appendix D).

\subsection*{Wave Motion}

When the linearized equation of motion
\begin{equation}
G^{linear}_{\mu \nu} + C^{linear}_{\mu \nu} = -8 \pi \thinspace G \,T_{\mu \nu}
\label{4.8}
\end{equation}
is decomposed according to ~(\ref{4.4}) and ~(\ref{4.5}), one finds that $C^{linear}_{00}$ vanishes, and
\begin{equation}
G^{linear}_{00} + C^{linear}_{00} = -\nabla^2 \varphi = -8 \pi G \, T_{00}. 
\label{4.9}
\end{equation}
The time-space components give
\begin{subequations}
\begin{equation}
G^{linear}_{0i} + C^{linear}_{0i} = \frac{1}{2} \nabla^2 \sigma^i_T - \partial_i \thinspace \dot{\varphi} +\frac{1}{4\mu} \thinspace \varepsilon^{imn} \thinspace \partial_m \nabla^2 \ \sigma^n_T= -8 \pi \, G \, T_{0i},
\label{4.10a}
\end{equation}
or
\begin{equation}
%\begin{array}{lll}
\frac{1}{2} \nabla^2 \sigma^i_T + \frac{1}{4 \mu} \thinspace \varepsilon^{imn} \ \partial_m \nabla^2 \sigma^n_T =  8 \pi G T^{0i}_T,
\label{4.10b}
\end{equation}
where
\begin{equation}
T^{0i}_T \equiv \left(\delta^{ij} - \frac{\partial_i \partial_j}{\nabla^2} \right) T^{0i}. \nonumber
%\end{array}
\end{equation}
\end{subequations}
In passing from ~(\ref{4.10a}) to ~(\ref{4.10b}), we used ~(\ref{4.9}) and conservation of $T^{\mu \nu}\negthickspace :  \partial_\mu T^{\mu \nu} =0$ (consistent with the linear approximation the ordinary, not covariant, divergence is involved).
For the space-space components we have
\begin{subequations}
\begin{equation}
\begin{array}{llll}
G^{linear}_{ij} + C^{linear}_{ij}= -\frac{1}{2} \left(\delta^{ij} - \frac{\partial_i \partial_j}{\nabla^2} \right) \lambda + \frac{1}{2} \  \partial_i  \dot{\sigma}^{j}_{T} + \frac{1}{2} \ \partial_j  \dot{\sigma}^i_T
 +\frac{1}{2} \Box h^{ij}_{TT} - \frac{\partial_i \partial_j}{\nabla^2} \ \ddot{\varphi} \\ [10pt]+ \frac{\varepsilon^{imn}}{4 \mu} \thinspace \partial_m \left(\partial_j \dot{\sigma}^n_T + \Box h^{jn}_{TT} \right)
 +\frac{\varepsilon^{jmn}}{4 \mu} \ \partial_m \left(\partial_i \dot{\sigma}^n_T + \Box h^{in}_{TT} \right)
= -8 \pi G T_{ij},
\end{array}
\label{4.11a}
\end{equation}
or
\begin{equation}
\frac{1}{2}  \Box  h^{ij}_{TT}   +   \frac{1}{4 \mu}   \varepsilon^{imn}  \partial_m   \Box \ h^{jn}_{TT}   +   \frac{1}{4 \mu}\  \varepsilon^{jmn}_\gamma  \partial_m \  \Box \ h^{in}_{TT} = -8  \pi  G \ T^{ij}_{TT},
\label{4.11b}
\end{equation}
with
\begin{eqnarray}
T^{ij}_{TT} \equiv T^{ij} - \frac{1}{2} \left(\delta_{ij} -\frac{\partial_i \partial_j}{\nabla^2} \right) \ T^{mm} 
+ \frac{1}{2} \left(\delta_{ij} + \frac{\partial_i \partial_j}{\nabla^2} \right) \frac{\partial_m \partial_n}{\nabla^2} \ T^{mn} 
- \frac{\partial_i \partial_m}{\nabla^2} \ T^{mj} \ -  \frac{\partial_j \partial_m}{\nabla^2} \ T^{mi}. \nonumber
\label{4.12}
\end{eqnarray}
\end{subequations}
Again, the previous equations ~(\ref{4.9}) and (4.10), were used together with conservation of $T^{\mu \nu}$ to pass from ~(\ref{4.11a}) to ~(\ref{4.11b}).

We see from ~(\ref{4.9}), ~(\ref{4.10b}) and ~(\ref{4.11b}) that the entire effect of the extension is to act on the left side of the Einstein equation -- the field side -- by spatial
derivative operators which carry the extension. These operators can be inverted and cast onto the right side -- the source side. So the entire effect of the extension is to
modify the source in its spatial dependence. In terms of the modified source, the equations regain their Einstein form, and the extension is invisible. From ~(\ref{4.9}), ~(\ref{4.10b})
and ~(\ref{4.11b}) we have
\begin{eqnarray}
\nabla^2 \varphi &=&  8 \pi G \ \widetilde{T}_{00}, \nonumber\\ [5pt]
\frac{1}{2} \  \nabla^2   \sigma^i_T &=&  \,\negthinspace 8 \pi G  \ \widetilde{T}^{0i}_{T}, \nonumber\\ [5pt]
\frac{1}{2} \  \Box \  h^{ij}_{TT} &=& \negthinspace \negthickspace \negthickspace -8 \pi G  \ \widetilde{T}^{ij}_{TT}.
\label{4.13}
\end{eqnarray}
The modified sources are given by
\begin{eqnarray}
\widetilde{T}^{00} &=& T^{00}, \nonumber\\ [5pt]
\widetilde{T}^{0i}_T &=& \frac{1}{1 + \frac{\nabla^2}{4 \mu^2}}  \left(\delta^{im} - \frac{1}{2 \mu} \ \varepsilon^{inm} \ \partial_n \right) T^{om}_T, \nonumber\\ [5pt]
\widetilde{T}^{ij}_{TT} &=& \frac{1}{1 + \frac{\nabla^2}{\mu^2}}  \ \bigg(\delta^{im} \ \delta^{jn} - \frac{1}{2 \mu} \ \varepsilon^{ikm} \ \delta^{jn} \ \partial_k 
 - \frac{1}{2 \mu} \ \varepsilon^{jkm} \ \delta^{in} \ \partial_k \bigg) \thinspace T^{mn}_{TT}.
\label{4.14}
\end{eqnarray}
Again we see that the extended theory supports gravitational waves with two polarizations $(h^{ij}_{TT})$ traveling with velocity of light.

The effect of the extension is felt when we examine the intensity of radiation.
In Einstein's theory, a monochromatic source whose energy momentum tensor possesses a definite frequency
\begin{equation}
T^{\mu \nu} (t, {\bf r}) = e^{-i \omega t} \ T^{\mu \nu} \ (\omega, {\bf r}) + c. c.,
\end{equation}
radiates power per unit angle in direction $\widehat{\bf k}$, \cite{rj4}
\begin{equation}
\frac{d P}{d \Omega} = \frac{G \omega^2}{\pi} \ \ T^{*ij}_{TT} \ (\omega, {\bf k})  \ \ T^{ij}_{TT} \ (\omega, {\bf k}),
\end{equation}
where
\begin{equation}
T^{\mu \nu} \ (\omega, {\bf k}) = \int d^3   r \ e^{-i {\bf k} \cdot {\bf r}} \ T^{\mu \nu} \ (\omega, {\bf r}).
\end{equation}
Let the direction of propagation be the third axis. Then $T^{ij}_{TT}$ takes the form
\begin{equation}
T^{ij}_{TT}\ (\omega, {\bf k}) = \left(
\begin{array}{ccc} 
T&S&0\\
S&-T&0\\
0&0&0
\end{array} \right),
\end{equation}
and definite polarizations correspond to $T \pm i \ S$.
Thus (4.15) becomes
\begin{equation}
\frac{d P}{d \Omega} = \frac{2G \omega^2}{\pi} \ \bigg(|T|^2 + |S|^2 \bigg).
\label{4.19}
\end{equation}
Each of the single polarizations $(T= \pm i S)$ contributes the same amount.\\
\begin{equation}
\frac{d P_\pm}{d \Omega} = \frac{4G \omega^2}{\pi} \ |T|^2
\end{equation}

For a similar computation in the extended theory, we need merely to replace $T^{ij}_{TT}$ with $\widetilde{T}^{ij}_{TT}$, which according to 
~(\ref{4.13}) takes the form
\begin{equation}
\widetilde{T}^{ij}_{TT} \ (\omega, {\bf k}) =\frac{1}{1-\frac{k^2}{\mu^2}}
\left(
\begin{array}{lr}
T-\ \frac{ik}{\mu} \ S&S + \ \frac{i k}{\mu} \ T\\ [5pt]
S + \ \frac{i k}{\mu} \ T&-T+\ \frac{ik}{\mu} \ S
\end{array} \right)
\label{4.20}
\end{equation}
Definite polarizations now correspond to $(T-\frac{i k S}{\mu}) \pm i (S + \frac{ikT}{\mu}) = (1 \mp \frac{k}{\mu}) ( T \pm i S)$, so that a single polarization still requires $T = \pm i S$. The radiated power
now reads
\begin{equation}
\frac{d P}{d \Omega} = \frac{2G \omega^2}{\pi (1-\frac{k^2}{\mu^2})^2} \left[ \bigg(1 + \frac{k^2}{\mu^2} \bigg) \bigg(|T|^2 + |S|^2 \bigg) + \frac{2ik}{\mu} \bigg(TS^* - S T^* \bigg) \right].
\end{equation}
For a single polarization $(T = \pm \ i S)$ we find
\begin{equation}
\frac{d P_\pm}{d \Omega}  \  = \ \frac{4G \omega^2}{\pi} \   |T|^2 \ \frac{1}{(1 \pm \frac{k}{\mu})^2}.
\end{equation}
Thus although each polarization travels with the velocity of light, they carry different intensities. The correction to the Einstein value grows with energy (owing to the triple derivatives in the extension), and for large
$\mu$ (negligible extension) it gives simply the suppression/enhancement factor $(1\mp 2\frac{k}{\mu})$. Thus the extension manifestly violates spatial reflection symmetry.

Note that there is no sign of the instability seen at small $k$ in Chern-Simons extended electromagnetism. Here however we encounter a singularity at $k =  \mu$, which reflects the fact that the spatial derivative
operator acting on $\Box h^{ij}_{TT}$ in ~(\ref{4.11b}), possesses zero modes, {\it i.e.} solutions to the homogenous equations exist. This allows for an arbitrary ``background" gravitational field to be present.
We find no further information about this in the linear theory; perhaps non-linear effects can clarify the situation.

\subsection*{Gravitational Energy-Momentum (Pseudo-) Tensor}

One way of identifying the energy-momentum (pseudo-) tensor of the gravitational field is to separate from the field tensor (``left" side) in the equation of motion its linear part and transfer the non-linear part to the source side (``right" side), combining it with
the matter energy-momentum tensor. Since the linear terms of the field tensor (on the left side) are conventionally conserved, so must be the entire right side. Also the expression is manifestly
symmetric in its space-time indices. In this way one constructs an ordinarily conserved, second rank symmetric (pseudo-) tensor, which is then identified with the total (matter + gravity) energy-momentum (pseudo-) tensor.

The above scheme can be carried out for the extended theory. We write ~(\ref{2.11}) as $G^{linear}_{\mu \nu} + C^{linear}_{\mu \nu} \ =  - \, 8 \pi \, G \, T_{\mu \nu}  - \Delta(G_{\mu \nu} + C_{\mu \nu})$ where $\Delta(G_{\mu \nu} + C_{\mu \nu}) = G_{\mu \nu}
+C_{\mu \nu} - G^{linear}_{\mu \nu} - C_{\mu \nu ~\cdot}^{linear}$ 
Evidently if we define $\tau_{\mu \nu} \ = \ T_{\mu \nu} + \frac{1}{8 \pi G} \ \Delta (G_{\mu \nu} + C_{\mu \nu})$
we have in $\tau_{\mu \nu}$ a conserved, symmetric energy-momentum (pseudo-) tensor. This, together with the velocity of light propagation of gravitational waves and the persistence of the Schwarzschild solution, shows that symmetry violation is effectively hidden in our extended gravity theory.

(In ref. \cite{rj6} there is a survey of alternative definitions for the energy-momentum (pseudo-) tensor of Einstein's gravity. Included is a construction based on Noether's theorem for translation
invariance, together with a Belinfante improvement.\ This yields a symmetric, conserved (pseudo-) tensor, which differs from the above by a superpotential. It would be interesting
to find also for the deformed theory the improved, symmetric Noether (pseudo-) tensor.)

\section{Conclusion}
Measuring the intensity of polarized gravity waves is now not yet feasable. So the modification of Einstein's theory that we have explored does not have an immediately apprehensible physical consequence. But the structure is interesting
in that diffeomorphism symmetry is broken in the action, yet the equations motion coincide with those of a modified, but symmetric theory. Specifically we find only two gravity wave
helicities, both propagating with the velocity of light. This is usually seen as a consequence of diffeomorphism invariance, but it persists
in our deformed theory.

The Stuckelberg mechanism for massive Abelian gauge fields provides an analogy to the situation with diffeomorphism symmetry encountered in our deformed gravity
model. The action for a massive $U(1)$ gauge field is not gauge invariant:
\begin{equation}
I_m = \int d^4 x \left(-\frac{1}{4} F^{\mu \nu} F_{\mu \nu} + \frac{1}{2} m^2 A^\mu A_\mu \right),
\label{5.1}
\end{equation}
\begin{eqnarray}
\delta_{\mbox{gauge}}\ A_\mu &=& \partial_\mu \lambda \nonumber\\
\delta_{\mbox{gauge}} \ I_m &=& \int d^4 x m^2 A^\mu \partial_\mu \lambda = -\int d^4 x m^2 \partial_\mu A^\mu \lambda.
\label{5.2}
\end{eqnarray}
Gauge symmetry is broken by $\partial_\mu A^\mu$. But the equation of motion
\begin{equation}
\partial_\mu F^{\mu \nu} + m^2 A^\nu = J^\nu
\label{5.3}
\end{equation}
enforces (for a conserved source current $J^\nu$) the vanishing of the gauge field divergence.
\begin{equation}
m^2 \partial_\nu A^\nu = 0
\label{5.4}
\end{equation}
So (\ref{5.3}) may also be presented in a gauge invariant form.
\begin{equation}
\partial_\mu F^{\mu \nu} + m^2 \left(g^{\mu \nu} - \frac{\partial^\mu \partial^\nu}{\Box}\right) A_\mu = J^\nu
\label{5.5} 
\end{equation}
But an important difference remains. If the mass term in the gauge field example is promoted to a dynamical ``field" variable $m^2 \to m^2 (x)$, then varying 
$m^2(x)$ [in analogy to varying $\theta (x)$ in our extended gravity theory] obtains
\begin{equation}
\frac{\delta I_m}{\delta m^2 (x)} = \frac{1}{2} A^\mu (x) \ A_\mu (x)
\label{5.6}
\end{equation}
But unlike in extended gravity, the resulting equation
\begin{equation}
A^\mu A_\mu = 0
\label{5.7}
\end{equation}
does not follow from the equation of motion (\ref{5.3}), obtained by varying $A_\mu$. Moreover, (\ref{5.7}) is an unacceptable equation since it eliminates the possiblility
of finding non trivial solutions to (\ref{5.3}) [One recognizes that the Higgs mechanism in unitary gauge provides kinetic and potential terms for the Higgs field $``m^2(x)"$,
and leads to an acceptable equation for that quantity.]

This research was performed at the Aspen Center for Physics and  was supported in part by funds from the U.S. Department of Energy, under cooperative research agreements
DE-FCO2-94-ER40818 and  DE-FG02-91-ER40676.

\section*{Added Note:}
S. Carroll, whom we thank, has informed us that cosmological consequences of the modified gravity theory (\ref{2.8}) taken in a sourceless, linear approximation, have been previously examined by  Lue {\it et al.} \cite{rj7} The only topic common
to both investigations is a study of gravity waves. The analysis in \cite{rj7} differs from ours in that their $\theta$ carries arbitrary time dependence, while we allow only linear dependence. Nevertheless,
physical conclusions are similar: \negthinspace polarized waves are suppressed/enhanced in the extended theory.
Also we thank D. Grumiller for remarks about the Kerr metric.

\appendixa

We derive (\ref{2.10}) from (\ref{2.9}). With (\ref{2.5}), $I_{CS}$ is given by
\begin{equation}
I_{CS} = -\frac{1}{2} \int d^4 x v_\mu K^\mu,
\end{equation}
where $K^\mu$ is defined (\ref{2.4}). Varying the connection and recalling the definition of the Riemann tensor (\ref{2.6}) shows that 
\begin{equation}
\delta I_{CS} = -\frac{1}{2} \int d^4 x v_\mu \varepsilon^{\mu \alpha \beta \gamma} R^\tau_{~\sigma \gamma \beta} \delta \Gamma^\sigma_{\alpha \tau}.
\label{A.2}
\end{equation}
But the variation of the connection arises by varying the metric tensor.
\begin{equation}
\delta \Gamma^\sigma_{\alpha \tau} = \frac{g^{\sigma \nu}}{2} (D_\alpha \delta g_{\nu \tau} + D_\tau \delta g_{\nu \alpha} - D_\nu g_{\alpha \tau})
\label{A.3}
\end{equation}
Continuing (\ref{A.2}) and (\ref{A.3}) results in
\begin{equation}
\delta I_{CS} = -\frac{1}{4} \int d^4 x v_\mu \varepsilon^{\mu \alpha \beta \gamma} R^{\tau \nu}_{~ \ \gamma \beta} (D_\alpha \delta g_{\nu \tau} + D_\tau \delta g_{\nu \alpha} - D_\nu \delta g_{\alpha \tau}).
\end{equation}
Because $R^{\tau \nu}_{~ \ \gamma \beta}$ is antisymmetric in $ [\tau, \nu]$, the first term in parentheses does not contribute and the remaining two combine. After a partial integration, we are left with
\begin{equation}
\delta I_{CS} = \frac{1}{2} \int d^4 x (v_\mu \varepsilon^{\mu \alpha \beta \gamma} D_\tau R^{\tau \nu}_{~\ \gamma \beta} + v_{\mu \tau} \varepsilon^{\mu \alpha \beta \gamma} R^{\tau \nu}_{~ \ \gamma \beta}) \delta g_{\nu \alpha}.
\label{A.5}
\end{equation}
In the first intergrand we use the Bianchi identity to replace $D_\tau R^{\tau \nu}_{~\gamma \beta}$ by $D_\gamma R^\nu_\beta - D_\beta R^\nu_\gamma$, while the second integral is rewritten in terms of the dual Riemann
tensor (\ref{2.7}).
\begin{equation}
\delta I_{CS} = \int \negthinspace d^4 x (v_\mu \varepsilon^{\mu \alpha \beta \gamma} D_\gamma R^\nu_\beta - v_{\mu \tau} {^*R^{\tau \nu \mu \alpha}}) \delta g_{\nu \alpha}
\label{A.6}
\end{equation}
Comparison of (\ref{A.6}) with (\ref{2.9}) establishes (\ref{2.10}).

\appendixb

We derive (\ref{2.12}) by computing explicity the covariant divergence of the Cotton tensor, whose formula we take from (\ref{2.7}), (\ref{2.9}) and (\ref{A.5}).
\begin{equation}
C^{\mu \nu} = - D_\tau \frac{v_\sigma}{2 \sqrt{-g}} \bigg({^*R^{\tau \mu \sigma \nu}} + {^*R^{\tau \nu \sigma \mu}} \bigg)
\end{equation}
Using the antisymmetry of ${^*R^{\tau \mu \sigma \nu}}$ in $[\tau, \mu]$, we can present $D_\mu C^{\mu \nu}$ as
\begin{eqnarray}
D_\mu C^{\mu \nu} =-D_\tau D_\mu \frac{v_\sigma}{2 \sqrt{-g}} {^*R^{\tau \nu \sigma \mu}}
+ [D_\tau, D_\mu] \frac{v_\sigma}{2 \sqrt{-g}} ({^*R^{\tau \nu \sigma \mu}} + \frac{1}{2} {^*R^{\tau \mu \sigma \nu}}).
\label{B.2}
\end{eqnarray}
The first contribution to $D_\mu C^{\mu \nu}$ vanishes. This is established by noting that there occurs
\begin{equation}
D_\mu \frac{v_\sigma}{2 \sqrt{-g}} {^*R^{\tau \nu \sigma \mu}} = \frac{v_{\mu \sigma}}{2 \sqrt{-g}} {^*R^{\tau \nu \sigma \mu}} 
+ \frac{v_{\sigma}}{2 \sqrt{-g}} \varepsilon^{\sigma \mu \alpha \beta} D_\mu R^{\tau \nu}_{~ \ \alpha \beta}. \nonumber
\end{equation}
Since ${^*R^{\tau \nu \sigma \mu}}$ is antisymmetric in $[\sigma, \mu]$ and $v_{\mu \sigma}$ is symmetric, the first term on the right is zero, and so is the second owing 
to the Bianchi identity satisfied by the Riemann tensor. The remainder of (\ref{B.2}) involves the commutator of covariant derivatives, and leads to
\begin{subequations}
\begin{eqnarray}
D_\mu C^{\mu \nu} =& \negthickspace \negthickspace   
\dfrac{v_\sigma}{2 \sqrt{-g}} \bigg[ \bigg({^*R^{\lambda \nu \sigma \mu}} + \dfrac{1}{2} {^*R^{\lambda \mu \sigma \nu}} \bigg) R^\tau_{~\lambda \mu \tau}
&\negthickspace \negthickspace \negthickspace + {^*R^{\tau \lambda \sigma \mu}} R^\nu_{~  \lambda\mu \tau} + \frac{1}{2} {^*R^{\tau \lambda \sigma \nu}} R^\mu_{~\lambda \mu \tau} \nonumber \\ 
&& \negthickspace \negthickspace \negthickspace + {^*R^{\tau \nu \sigma \lambda}} R^\mu_{~ \lambda \mu \tau} + \frac{1}{2} {^*R^{\tau \mu \sigma \lambda}} R^\nu_{~\lambda \mu \tau} \bigg] \nonumber \\ 
=&  \negthickspace \negthickspace \dfrac{v_\sigma}{2 \sqrt{-g}} \bigg[ - \bigg({^*R^{\lambda \nu \sigma \mu}} + \dfrac{1}{2} {^*R^{\lambda \mu \sigma \nu}} \bigg) R_{\lambda \mu} 
& \negthickspace \negthickspace \negthickspace +\bigg({^*R^{\tau \nu \sigma \lambda}} + \frac{1}{2} {^*R^{\tau \lambda \sigma \nu}} \bigg) R_{\lambda \tau} \nonumber \\
&& \negthickspace \negthickspace \negthickspace +\bigg({^*R^{\tau \lambda \sigma \mu}} + \frac{1}{2} {^*R^{\tau \mu \sigma \lambda}} \bigg) R^\nu_{~ \lambda \mu \tau} \bigg].
\end{eqnarray}
The quantities involving the Ricci tensor vanish owing to its symmetry. The last term in brackets is expanded by using the antisymmetry of ${^*R^{\tau \lambda \sigma \mu}}$ in $[\tau, \lambda]$.
Thus we are left with
\begin{eqnarray}
D_\mu C^{\mu \nu} &=& \frac{v_\sigma}{4 \sqrt{-g}} \bigg[{^*R^{\tau \lambda \sigma \mu}} \bigg(R^\nu_{~\lambda \mu \tau}- R^\nu_{\tau \mu \lambda} \bigg)
+{^*R^{\tau \mu \sigma \lambda}} R^\nu_{~\lambda \mu \tau} \bigg] \nonumber \\ 
&=& \frac{v_\sigma}{4 \sqrt{-g}} \bigg[{^*R^{\tau \lambda \sigma \mu}} R^\nu_{~\mu \lambda \tau} + {^*R^{\tau \mu \sigma \lambda}} R^\nu_{~\lambda \mu \tau} \bigg] \nonumber \\ 
&=& \frac{v_\sigma}{2 \sqrt{-g}} {^*R^{\tau \lambda \sigma \mu}} R_{\lambda \tau ~ \mu}^{~\ \nu}.
\end{eqnarray}
\end{subequations}
 Cyclic properties of the Riemann tensor allow passage from one expression to the next in (B.3b). Finally we use the identity
\begin{equation}
{^*R_{~\lambda}^{\tau ~\sigma \mu}} R^\lambda_{~\tau \nu \mu} = \frac{1}{4} \delta^\sigma_\nu {^*RR},
\end{equation}
to conclude that
\begin{equation}
D_\mu C^{\mu \nu} = \frac{v^\nu}{8 \sqrt{-g}} {^*RR},
\end{equation}
in agreement with (\ref{2.12}).

\appendixc

We establish the vanishing of the last term in (\ref{3.8}). (The calculation is entirely three dimensional, so we omit the superscript ``3".)
Note first that according to the definition (\ref{3.4})
\begin{eqnarray}
\varepsilon^{mij} D_iN^n_j &=& \varepsilon^{mij} \bigg(\partial_i \frac{1}{\sqrt{N}} \bigg) \bigg(D_j D^n \sqrt{N} \bigg)
+ \varepsilon^{mij} \frac{1}{\sqrt{N}} D_i D_j D^n \sqrt{N} \nonumber \\
&=& \varepsilon^{mij} \bigg(-\frac{\partial_i \sqrt{N}}{\sqrt{N}} N^n_j + \frac{1}{2 \sqrt{N}} [D_i, D_j] D^n \sqrt{N} \bigg).
\label{C.1}
\end{eqnarray}
The action of the commutator of covariant derivatives on $D^n \sqrt{N}$ produces \\$-\frac{1}{2} \varepsilon^{mij} \frac{1}{\sqrt{N}} R^{nk}_{~ \ ij} \partial_k \sqrt{N} = \varepsilon^{nk \ell} G^m_\ell
\frac{\partial_k \sqrt{N}}{\sqrt{N}}$, where we have used (\ref{3.3b}). Thus (\ref{C.1}) becomes
\begin{equation}
\varepsilon^{mij} D_i N^n_j = -\partial_i \ell n \sqrt{N} \bigg(\varepsilon^{mij} N^n_j - \varepsilon^{nij} G^m_j \bigg).
\end{equation}
Using this equality in (\ref{3.8}) establishes the desired result.

\appendixd
We calculate ${^*RR}$ -- a measure of symmetry breaking in our Chern-Simons extended gravity theory -- to quadratic order in $h^{\mu \nu}$. The Riemann tensor in linear order is
\begin{equation}
R^{linear}_{\sigma \tau \alpha \beta} = \frac{1}{2} \bigg(\partial_\beta \partial_\tau h_{\sigma \alpha}- \partial_\alpha \partial_\tau h_{\sigma \beta}
+ \partial_\alpha \partial_\sigma h_{\tau \beta} - \partial_\beta \partial_\sigma h_{r \alpha} \bigg).
\end{equation}
The dual reads
\begin{equation}
{^*R_{\sigma \tau}}^{~\mu \nu ~ linear} = \frac{1}{2} \varepsilon^{\mu \nu \alpha \beta} \partial_\alpha \bigg(\partial_\sigma h_{\tau \beta}
-  \partial_\tau h_{\sigma \beta} \bigg).
\end{equation}
Therefore we have
\begin{equation}
^*RR^{quadratic} = \varepsilon^{\mu \nu \alpha \beta} \partial_\alpha \bigg(\partial_\sigma h_{\tau \beta}
- \partial_\tau h_{\sigma \beta} \bigg) \partial_\nu \partial^\sigma h^\tau_\mu.
\end{equation}
For plane monochromatic waves, $\partial_\alpha \partial_\sigma h_{\tau \beta} = - k_\sigma k_\alpha h_{\tau \beta}, ( k^\gamma k_\gamma =0)$, and the above becomes
\begin{equation}
^*RR^{quadratic} =\varepsilon^{\mu \nu \alpha \beta} k_\alpha k_\nu (k^\gamma k_\gamma h^\tau_{\mu} h_{\tau \beta} - k^\sigma h_{\sigma \beta} k^\tau h_{\tau \mu})
\end{equation}
This vanishes for a variety of reasons.

\end{document}